\documentclass[3p,twocolumn]{elsarticle}

\usepackage{lineno}
\usepackage[fleqn]{amsmath}
\usepackage{amssymb}
\usepackage{bm}
\usepackage{physics}
\usepackage{subcaption}
\usepackage{graphicx}
\usepackage{color}
\usepackage{import}
\graphicspath{{images/}, {images/ex1/}, {images/ex2/}}
\usepackage{booktabs} 
\journal{}

\makeatletter
\def\ps@pprintTitle{%
  \let\@oddhead\@empty
  \let\@evenhead\@empty
  \def\@oddfoot{\reset@font\hfil\thepage\hfil}
  \let\@evenfoot\@oddfoot
}
\makeatother

\begin{document}

\begin{frontmatter}

\title{Topology optimization of wind farm layouts}
\date{}

\author[mymainaddress]{Nicol\`{o} Pollini\corref{mycorrespondingauthor}}
\cortext[mycorrespondingauthor]{Corresponding author}
\ead{nicolo.pollini@mail.polimi.it}

\address[mymainaddress]{Technical  University  of  Denmark (DTU), DTU Wind Energy, Frederiksborgvej 399, 4000 Roskilde, Denmark}

\begin{abstract}
A novel approach for the solution of the wind farm layout optimization problem is presented.
The annual energy production is maximized with constraints on the minimum and maximum number of wind turbines placed, and on the minimum spacing between the wind turbines.
The proposed approach relies on a density-based topology optimization method, where continuous density variables varying between zero and one are assigned to each potential wind turbine location.
A wind turbine exists if its associated variable equals one, otherwise it does not exist if the associated variable is zero.
Intermediate values of the density variables are penalized with interpolation schemes traditionally used in the context of multi-material structural topology optimization.
The penalized intermediate values of the design variables become uneconomical and the optimization algorithm is implicitly pushed towards a preference of crisp $0$-$1$ final values.
The optimization problem is solved with a gradient-based algorithm based on first-order information.
Because of the proposed problem formulation, the functions involved are formulated explicitly in terms of the design variables and their analytical gradients can be calculated directly.
The numerical results highlight the capability of the proposed approach in finding far from intuitive wind farm layouts with small computational resources and time.
\end{abstract}

\begin{keyword}
topology optimization \sep wind farm \sep layout optimization \sep interpolation scheme
\end{keyword}

\end{frontmatter}

\section{Introduction}
\label{sec:intro}

The pressure to further develop clean technologies for energy production with minimized carbon emissions is significantly increasing \cite{bloomberg2020new}.
Wind energy is a broadly available resource and has extraordinary low life-cycle pollutant emissions.
In fact, wind energy technologies have the potential for being the driving force in the global green energy transition \cite{veers2019grand}.
The continuous search for better solutions to reduce costs, has led to the installation of clusters of wind turbines with the purpose of maximizing energy production while reducing costs associated e.g. to installation and maintenance.
These clusters of wind turbines are more commonly referred to as wind farms in the literature.
The design of a wind farm is a highly complex and multidisciplinary process.
When a wind turbine extracts power from the wind, it generates a wake of turbulence that propagates downwind, so that the wind speed and thus the power extracted by the downstream wind turbines is reduced \cite{samorani2013wind}.
One of the major challenges in the design of a wind farm is the definition of the best layout of wind turbines so that the wake effects between adjacent wind turbines are minimized and the power extracted from the wind farm is maximized.
The wind farm layout optimization problem consists in finding the optimal positioning of a certain number of wind turbines within a wind farm while minimizing (or maximizing, depending on the case) one or several objectives, and while satisfying certain constraints.

Since the seminal work of Mosetti et al. in $1994$ \cite{mosetti1994optimization}, over the last $25$ years the wind farm layout optimization problem has received a growing attention \cite{tesauro2012state,herbert2014review}.
Most of the methods in the literature use gradient-free heuristics (\cite{khan2013iterative} and references therein), such as genetic algorithms \cite{mosetti1994optimization,grady2005placement,elkinton2008optimizing,gonzalez2013new,gao2014study,gao2015wind,gao2016optimization,mytilinou2019techno}, 
the random search algorithm \cite{feng2015solving,feng2017design,feng2017wind}, a simulated annealing algorithm \cite{rivas2009solving}, evolutionary algorithms \cite{kusiak2010design,reddy2020wind}, and the coral reef algorithm \cite{salcedo2014offshore}.
Even though these methods have proven effective for certain sizes of the wind farm layout optimization problem, they are expected to become less effective for large problems. 
As it is discussed also by Stanley and Ning \cite{stanley2019massive}, with the increase of the number of wind turbines and consequently of the design variables, the problem size increases significantly because of its combinatorial nature. As a consequence, the computational cost required by gradient-free optimization methods may become prohibitive jeopardizing the adoption of optimization-based wind farm design methods in the engineering practice.
In this work, we rely on a gradient-based approach to solve the wind farm layout optimization problem.

In the literature, some authors relied on gradient-based optimization approaches to reduce the computational cost of the wind farm layout optimization problem. 
For example, Thomas and Ning in \cite{thomas2018method}  propose a method for reducing multi-modality in the wind farm layout optimization problem, namely the Wake Expansion Continuation (WEC) method. They combine it with a gradient-based optimization approach and compare the results with those obtained with a gradient-free method and a gradient-based method without WEC. 
In \cite{stanley2019massive} Stanley and Ning present a boundary-grid parametrization. This parametrization allows to reduce the number of design variables to only a few even for design cases involving hundreds of wind turbines. The layout optimization problem is solved with a gradient-based algorithm based on sequential quadratic programming. 
The results show that the performance of the final design is comparable to those achieved optimizing the position of each turbine individually.
In \cite{baker2019best} Baker et al. present the results of two case studies regarding the wind farm layout optimization problem. 
They collect the results submitted by several members of the optimization and wind energy communities, for two wind farm layout design cases: the first based on a given wake model and an optimization approach of choice; the second based on a wake model and an optimization approach both selected by the participants. 
In the first design case, it emerges from the results submitted by the participants that gradient-based methods converge to final layouts with higher optimized annual energy production, for the same given number of wind turbines and optimization constraints.

The majority of the studies related to the gradient-based optimization of wind farm layouts assume a fixed number of wind turbines and search for their optimized placement.
The case with a variable number of wind turbines is more complex because it involves discrete variables in the problem formulation, and it has received less attention in the literature.
The optimization of wind farm layouts with variable number of wind turbines to be placed has been traditionally studied with gradient-free methods. 
Mosetti et al. in $1994$ \cite{mosetti1994optimization} first addressed this type of problem. They divided the wind farm domain into $100$ cells where turbines could be placed, and then used a genetic algorithm to optimize the wind turbine number and placement within the given grid. 
Since then, several authors used different optimization algorithms improving the solutions achieved (e.g. \cite{grady2005placement,ituarte2011optimization,moorthy2015new,fischetti2016proximity,zergane2018optimization, stanley2021objective}).
To the best of the author's knowledge, none of the contributions in the literature has discussed the gradient-based wind farm layout optimization for a variable number of wind turbines to be placed. 

Thus, in this paper we present a novel approach for the solution of the wind farm layout optimization problem where a minimum and maximum number of turbines to be placed is defined.
We rely on a density-based topology optimization approach, where continuous density variables that vary between zero and one are assigned to each potential wind turbine location. 
We consider a reference grid of potential wind turbine locations with given coordinates. To each potential wind turbine location in the reference grid belongs a design variable representing a fictitious wind turbine density: the density is one if a turbine exists at that location in the grid, otherwise it is zero. 
In general, topology optimization \cite{bendsoe2003topology} is a computational method aimed at optimizing the distribution of one or several materials in a given continuum design domain. Since the seminal paper of Bends{\o}e and Kikuchi in 1988 \cite{bendsoe1988generating}, topology optimization has undergone a tremendous development in several directions and fields \cite{sigmund2013topology,deaton2014survey}. 
In the design of load bearing structures, topology optimization is strongly connected to finite element analysis where the optimized topological layout depends on whether the optimizer places solid material (i.e. density value of $1$) or void (i.e. density value of $0$) in each finite element. 
They key for solving the resulting binary problem is the use of material interpolation functions. 
The binary variables are treated as continuous variables in the interval $[0,1]$, and the intermediate values between $0$ and $1$ are penalized by means of the material interpolation functions. 
This translates to finite elements with penalized low stiffness that produce uneconomical designs, from an optimization perspective. 
In this way, the optimization algorithm is implicitly pushed towards a preference of final $0$-$1$ designs even within a continuous problem formulation. 
In this work we will also rely on interpolation functions to converge towards discrete final wind farm layouts from a continuous problem definition.
We will consider the Annual Energy Production as the objective function maximized, with constraints on the minimum and maximum number of wind turbines allocated, and with a constraint on the minimum distance between neighboring allocated wind turbines.
The optimization is performed with a gradient-based approach relying on the analytical gradients of the functions involved.
The parametrization of the problem allows to eliminate the dependence of the wake losses on the variables of the problem. This means that the wind speed deficits for a given wind farm grid are precomputed before the optimization process is started. This allows also to replace the specific wake model selected, without affecting the computation of the gradients and the optimization algorithm.
The numerical results show that the proposed approach successfully identifies non-intuitive optimized wind farm layouts with small computational resources and time.

The reminder of the article is organized as follows: in Sec.~\ref{sec:problem} we present the problem formulation with details on the design variables of the problem, and the objective and constraint functions; in Sec.~\ref{sec:optappr} we provide important details on the interpolation functions adopted, on the optimization problem formulation, and on additional computation considerations relevant for the solution of the problem at hand; numerical results are presented and discussed in Sec.~\ref{sec:examples}. Sec.~\ref{sec:end} concludes the paper with final considerations.

\section{Problem formulation}
\label{sec:problem}
In this section we present the formulation of the optimization problem at hand.
In particular we provide the details of the design variables, and of the objective and constraints functions involved.

\subsection{Design variables}
\label{sec:vars}
We consider a predefined grid of potential wind turbines. 
The grid consists of $N$ points with coordinates $(x_i,y_i)$, for $i=1,...,N$. 
To each $i$-th point on the grid there is an associated binary design variable $\rho_i$, for for $i=1,...,N$:
\begin{equation}\label{eq:binvars}
    \rho_i =
    \begin{cases}
    1, & \text{ if turbine $i$ exists at location $i$} \\
    0, & \text{ if turbine $i$ doesn't exist at location $i$}
    \end{cases}
\end{equation}
The design variables $\rho_i$ are collected in the vector $\bm{\rho}$.

\subsection{Objective function}
\label{sec:objfunc}
In the literature, different cost models have been used to for the wind farm layout problem. 
The review article by Herbert-Acero et al. \cite{herbert2014review} lists several cost models present in the literature. 
Among those models, the Annual Energy Production is one of the most common cost formulations considered (e.g. \cite{kusiak2010design,chowdhury2013optimizing,gebraad2017maximization,quick2017optimization}). 
Thus, in this work the objective function maximized is the annual energy production of the wind farm.
Formally, we define the optimization problem as a minimization one, and the objective function minimized is:
\begin{equation}\label{eq:aep}
    f(\bm{\rho}) = - AEP(\bm{\rho})
\end{equation}
In Eq.~\eqref{eq:aep} $AEP$ is the annual energy production, which is defined as follows:
\begin{equation}
    AEP(\bm{\rho}) = \left( \sum_{i=1}^{N_{wd}} f_i \, P_{f}^i \right) 8760 \frac{hrs}{year}
\end{equation}
where $N_{wd}$ is the number of wind directions considered, $f_i$ is the frequency of the considered wind speed in direction $i$, and $P_{f,i}$ is the wind farm power production for incoming wind in direction $i$.
The total wind farm power $P_{f,i}$ for each wind direction considered is calculated as follows:
\begin{equation}\label{eq:power_f}
    P_{f}^i = \sum_{j=1}^{N} \rho_j \, P_{wt,j}^i
\end{equation}
with 
\begin{equation}\label{eq:power_t}
    P_{wt,j}^i =
    \begin{cases}
    0         & V_{e,j}^i < V_{ci} \\
    P_{r}\left(\frac{V_{e,j}^i - V_{ci}}{V_{r}-V_{ci}} \right)^3 & V_{ci} \leq V_{e,j}^i < V_{r}\\
    P_{r} & V_{r} \leq V_{e,j}^i < V_{co}\\
    0         & V_{e,j}^i > V_{co}
    \end{cases}
\end{equation}
where $V_{e,j}^i$ is the effective wind speed at turbine $j$ for the wind direction $i$, $V_{ci}$ is the cut-in wind speed, $V_{co}$ is the cut-out wind speed, $V_{cr}$ is the rated wind speed, and $P_{r}$ is the wind turbine rated power.
The effective wind speed $V_{e,j}^i$ is defined as follows:
\begin{equation}\label{eq:totloss}
\begin{split}
 V_{e,j}^i &= V_{\infty} \left[1 - \left( \frac{\Delta V}{V_{\infty}} \right)_{tot,j}^i \right] \\ 
 \left( \frac{\Delta V}{V_{\infty}} \right)_{tot,j}^i &= \sqrt{\sum_{k\in\mathcal{S}_j^i} \rho_k \left( \frac{\Delta V}{V_{\infty}} \right)_k^2}
\end{split}
\end{equation}
In Eq. \eqref{eq:totloss}, $V_{\infty}$ is the undisturbed wind speed, and for the $i$-th wind turbine and $j$-th wind direction $\left( \frac{\Delta V}{V_{\infty}} \right)_{toot,j}^i$ is the total wind speed loss due to the wakes generated by the upstream wind turbines, $\mathcal{S}_j^i$ is the set of indices of upstream wind turbines  each causing a wind speed loss equal to $\left( \frac{\Delta V}{V_{\infty}} \right)_k$, for $k \in \mathcal{S}_j^i$.

Different wake models are available in the literature to calculate the total wind speed deficit for each turbine in a given wind farm \cite{shakoor2016wake,goccmen2016wind,archer2018review}.
In this work, the wind speed deficit is calculated as in \cite{stanley2019massive} 
with a modified version of the 2016 Bastankhah Gaussian wake model \cite{bastankhah2016experimental} proposed in \cite{thomas2018method}, which is defined as follows:
\begin{equation}\label{eq:wakemod}
\begin{split}
  \frac{\Delta V}{V_{\infty}} = & \left(1- \sqrt{1- \frac{C_T \cos \gamma}{8\sigma_y \sigma_z / d^2}} \right)  \exp \left( -0.5 \left( \frac{y-\delta}{\sigma_y}\right)^2\right) \\
& \exp \left( -0.5 \left( \frac{z-z_h}{\sigma_z}\right)^2\right)
\end{split}
\end{equation}
where $\frac{\Delta V}{V_{\infty}}$ is the velocity deficit in the wake, $C_T$ is the thrust coefficient, $\gamma$ is the yaw angle assumed to be zero in this work (i.e. $\cos \gamma = 1$), $y-\delta$ and $z-z_h$ are the cross-stream horizontal and vertical distances from the wake center and the point of interest, $\sigma_y$ and $\sigma_z$ are the standard deviations of the wake deficit in the cross-stream horizontal and vertical directions. 
These standard deviations are defined as:
\begin{equation}\label{eq:wakestd}
\begin{split}
& \sigma_y = k_y(x-x_0) + \frac{D \cos \gamma}{\sqrt{8}}\\
& \sigma_z = k_z(x-x_0) + \frac{D }{\sqrt{8}}
\end{split}
\end{equation}
where $D$ is the diameter of the wind turbine generating the wake, $x-x_0$ is the downstream distance between two turbines, and lastly $k_y$ and $k_z$ are unitless parameters that depend on the free-stream turbulence intensity:
\begin{equation}
    k_y, k_z = 0.3837\, TI + 0.003678
\end{equation}
In this work $\sigma_y=\sigma_z$ because $\gamma=0$. 
Moreover, for the numerical examples we consider a value of the parameter $TI$ equal to $0.075$.

\subsection{Constraint functions}
\label{sec:constfunc}
The final layout results from the placement of minimum $N_{min}$ and maximum $N_{max}$ wind turbines.
$N_{min}$ and $N_{max}$ are predefined values.
These requirements can be formulated through two linear inequality volume constraints as follows:
\begin{equation}\label{eq:Vmin}
\begin{split}
   & g_{minV} = -\frac{\sum_{i=1}^{N} \rho_i}{N}  + V_{min}^* \leq  0\\ 
   & \text{ with: } V_{min}^*= N_{min} / N
\end{split}
\end{equation}
\begin{equation}\label{eq:Vmax}
\begin{split}
   & g_{maxV} = \frac{\sum_{i=1}^{N} \rho_i}{N} -  V_{max}^* \leq 0\\
   & \text{ with: } V_{max}^*= N_{max} / N
\end{split}
\end{equation}

We impose $N(N$-$1)$ additional constraints on the spacing between the wind turbines. In particular, we impose a distance between the wind turbines greater than or equal to two rotor diameters. Thus, for each wind turbine $i$ we have the following $N$-$1$ local volume constraints:
\begin{equation}\label{eq:locvolcon1}
g_{locV,ij} = \rho_i + \rho_j  - 1 \leq 0, \text{ for } i,j \in \mathcal{N}_{i}
\end{equation}
where $\mathcal{N}_{i}$ is a set containing the indices of adjacent wind turbines around the $i$-th wind turbine, and it is defined as follows:
\begin{equation}
\begin{split}
\mathcal{N}_{i} = \bigg\{ & i,j \mid \sqrt{(x_i-x_j)^2 + (y_i-y_j)^2} \leq 2 D; \\
& i \neq j; \, i,j \in \{1,\dots,N\} \bigg\}
\end{split}
\end{equation}
where $D$ is the wind turbine rotor diameter. 
In matrix form Eq. \eqref{eq:locvolcon1} reads:
\begin{equation}\label{eq:locvolcon2}
\textbf{g}_{locV} = \textbf{H} \, \bm{\rho}  - \textbf{1} \leq 0
\end{equation}
In Eq. \eqref{eq:locvolcon2}, $\textbf{H}$ is a $[N(N$-$1) \times N]$ sparse matrix. 
Each row has two non-zero entries, and they are:
\begin{equation}
H_{i,i} = 1, \, H_{i,j} = w_{i,j}
\end{equation}
with
\begin{equation}
w_{i,j}=
\begin{cases}
    1, & i,j \in \mathcal{N}_{i} \\
    0, & \text{otherwise}
    \end{cases}
\end{equation}

\subsection{Initial optimization problem formulation}
\label{sec:initoptprob}
We can now present the initial problem formulation in its mixed-integer nonlinear form:
\begin{equation}\label{eq:initoptprob}
\begin{split}
\underset{\bm{\rho} \, \in \, \mathbb{Z} ^{N}}{\text{minimize: }} & f(\bm{\rho})\\
\text{subject to: } & g_{\text{minV}}(\bm{\rho}) \leq 0\\
 & g_{\text{maxV}}(\bm{\rho}) \leq 0\\
 & g_{\text{locV,}ij}(\bm{\rho}) \leq 0,  \text{ for } i,j \in \mathcal{N}_{i}\\
 &  \rho_i \in \{0,\; 1\},  \text{ for } i=1,\dots,N\\
\end{split}
\end{equation}
This problem can be solved, for example, by means of gradient-free methods, such as genetic algorithms. 
In the following section, we discuss the details of the continuous relaxation of problem \eqref{eq:initoptprob}.
A major advantage of the approach discussed herein is that the wind speed deficits across the wind farm are computed only once before the optimization analysis begins, for the given wind farm layout grid and wake model considered. Thus, the specific wake model adopted does not depend on the design variables of the problem, and as a consequence it does not affect the complexity of the functions' gradients, which in this approach are quite simple to compute analytically. 

\section{Optimization approach}
\label{sec:optappr}
In this section, we relax the definition of problem \eqref{eq:initoptprob} using only continuous variables. 
We rely on interpolation functions in the optimization problem re-formulation in order to reach final discrete solutions from a strictly continuous formulation. 
The role of the interpolation functions essentially is to penalize the intermediate values of the density variables $\rho_i$ making them uneconomical.
As a result, the optimization algorithm is implicitly pushed towards a preference of crisp $0$-$1$ final density values, thus promoting the convergence to discrete solutions. 

\subsection{Material interpolation techniques}
\label{sec:matint}
In the proposed approach, the definition of the design variables \eqref{eq:binvars} is relaxed, such that the variables can assume all the values between zero and one: $\rho_i\in [0,1]$.
In this way, the optimization problem at hand is continuous, and efficient gradient-based optimization algorithm can be used. 
The continuous variables $\rho_i$ are here referred to as densities. 
It is important to note that they do not model an actual material density, but they represent a fictitious wind turbine density in the given wind farm.

A practical discrete solution starting from a continuous formulation is achieved through the application of material interpolation techniques well-established in the field of topology optimization. 
The most commonly used interpolation is the SIMP model (Solid Isotropic Material with Penalization, \cite{bendsoe1989optimal}) which has proven to be  successful for a large number of applications \cite{sigmund2013topology}.
For more details on SIMP and several other interpolations techniques, the interested reader is referred to \cite{bendsoe1999material}.
In the current study, we performed numerical experiments utilizing either SIMP or the so-called RAMP (Rational Approximation of Material Properties)
interpolation function \cite{stolpe2001alternative}. 
The latter was chosen for the final problem formulation, as it proved to be more effective and promising in achieving final discrete solutions.

\begin{figure}[htbp!]
\centering
\includegraphics[trim=0 0 0 0,clip,width=.9\linewidth]{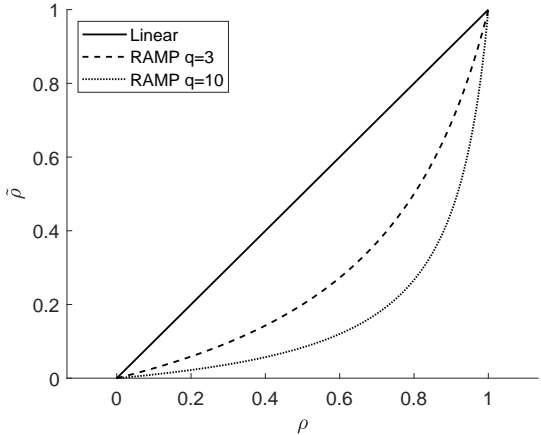}
\caption{Examples of interpolation with the RAMP model for different values of the penalty parameter $q$}
\label{fig:ramp}
\end{figure}

Thus, we interpolate each density variable $\rho_i$ for $i=1, \dots, N$ by means of the RAMP model as follows:
\begin{equation}\label{eq:rhoRAMP}
\tilde{\rho}_i = \frac{\rho_i}{1+q(1-\rho_i)}, \quad 0\leq \rho_i \leq 1
\end{equation}
For $q=0$ the interpolation is linear, and there is no penalization of the intermediate values, see Fig. \ref{fig:ramp}. 
For increasing values of the penalty parameter $q$, the intermediate values of the density variables correspond to vanishing wind turbines in the layout and in this way the algorithm is pushed towards the preferences of zero if the turbine is not desired in the final layout, or one if the turbine is desired by the algorithm in the final layout.

To the best of the author's knowledge, the current contribution represents the ﬁrst attempt to apply material interpolation techniques in the context of wind farm layout optimization.

\subsection{Final optimization problem formulation}
\label{sec:endoptprob}
Finally, we can present the re-formulation of the optimization problem at hand:
\begin{equation}\label{eq:endoptprob}
\begin{split}
\underset{\bm{\rho} \, \in \, \mathbb{R} ^{N}}{\text{minimize: }} & f(\tilde{\bm{\rho}})\\
\text{subject to: } & g_{\text{minV}}(\bm{\rho}) \leq 0\\
 & g_{\text{maxV}}(\bm{\rho}) \leq 0\\
 & g_{\text{locV,}ij}(\bm{\rho}) \leq 0,  \text{ for } i,j \in \mathcal{N}_{i}\\
 &  0 \leq \rho_i \leq 1, \text{ for } i=1,\dots,N\\
 \text{with: } &  \tilde{\rho}_i = \frac{ \rho_i}{1+q(1- \rho_i)}, \text{ for } i=1,\dots,N\\
\end{split}
\end{equation}
The topology optimization problem of Eq.~\eqref{eq:endoptprob} has been solved with an iterative gradient-based optimization algorithm, namely the Method of Moving  Asymptotes (MMA) \cite{svanberg1987method}. This algorithm relies on first order  information. Therefore, the gradients of the objective function and of the constraint need to be calculated. The constraint functions depend linearly on the design variables. The dependency of the objective function on the design variables is explicit. Hence, the gradients of the functions involved in the problem formulation can be directly calculated.

\subsection{Computational considerations}
\label{sec:compcons}
To successfully adopt the MMA algorithm for the nonlinear and nonconvex optimization problem at hand defined in Eq.~\eqref{eq:endoptprob}, practical and conservative measures are included in the implementation of the optimization approach proposed here. These include a continuation scheme for the control of the values attained by the penalty parameter; moving limits; and convergence criteria.

It should be noted that the optimization problem \eqref{eq:endoptprob} is nonlinear, nonconvex and it may converge to non-unique final solutions. 
The design solutions obtained are in general local minima of the problem, and they depend on the starting point of the optimization analysis, and on the particular continuation scheme adopted. 
Nevertheless, based on the author's experience, the continuation scheme discussed below gives a good balance between ease of convergence of the optimization analysis, and performance of the optimized solution.

\subsubsection{Continuation scheme for parameter control}
\label{sec:contscheme}
The optimization problem stated in Eq.~\eqref{eq:endoptprob} includes several nonlinear components, namely, the RAMP interpolation functions (Eq.~\eqref{eq:rhoRAMP}). Therefore, difficulties to converge smoothly towards a good optimized solution are expected. A common approach in the field of topology optimization for promoting a smooth convergence of the optimization analysis is to increase gradually the parameters that control the degree of nonlinearity. In our case this applies to the parameter $q$.
The parameter $q$ during the optimization analysis is increased from a minimum value to a maximum value by predefined steps: $q=[q_{min}:\Delta q:q_{max}]$.
Additionally, a conservative moving limit strategy is applied during the optimization process. In each optimization iteration $i$ the updates of $\bm{\rho}$ are searched in the vicinity of the solution of the previous iteration $i-1$: $\bm{\rho}_{i-1} - \textbf{m}_l \leq \bm{\rho}_i \leq \bm{\rho}_{i-1} + \textbf{m}_l$. Specific details regarding the values of these parameters are given in the numerical examples of Section \ref{sec:examples}.

\subsubsection{Convergence criteria}
\label{sec:convcrit}
The optimization analysis is assumed to have reached the final solution in a $i$-th iteration after a maximum of $i_{max}$ iterations, or once we have that $ \Delta \bm{\rho} < d$, where $\Delta \bm{\rho} = \norm{ \bm{\rho}_i - \bm{\rho}_{i-1}  }$, and $q=\Bar{q}$. The values of $i_{max}$, $d$, and $\Bar{q}$ are given in Section \ref{sec:examples}.


\section{Numerical examples}
\label{sec:examples}
In this section, we discuss the results obtained in two numerical examples. 
These examples are inspired by the ones considered in the Case Study 1 of \cite{baker2019best}.  
In particular, the first example concerns the optimization of a wind farm layout with a circular shape of radius $1300$ m.
We consider a wind farm reference grid with a spacing of $200$ m between the potential wind turbines.
In this case we require a minimum of $16$ and a maximum of $64$ wind turbines to be placed.
The second example concerns the optimization of a wind farm layout with a circular shape of radius $3000$ m.
We consider also in this case a reference grid with a spacing of $200$ m between the available wind turbines.
In this case, we require a minimum of $64$ and a maximum of $256$ wind turbines to be placed.

The wind farm analysis and optimization codes have been implemented by the author in MATLAB ver. 9.8.0. 
As it has been already mentioned, the optimization problem \eqref{eq:endoptprob} has been solved with the MMA MATLAB implementation kindly provided by Prof. Krister Svanberg.
For comparison, the optimization problem \eqref{eq:endoptprob} has been solved also using the Sequential Quadratic Programming (SQP) algorithm implemented in the \texttt{fmincon} function, which is part of the MATLAB Optimization Toolbox ver. 8.5.
Additionally, the optimization problem \eqref{eq:initoptprob} has been solved with the Genetic Algorithm (GA) available in the MATLAB Global Optimization Toolbox ver. 4.3. 
For MMA the following settings were used:
$q_{min} = 0$;
$\Delta q =0.5$ every $10$ iterations; 
$q_{max} = 10$;  
$m_{l} = 0.1$; 
$i_{max} = 1000$; 
$d = 1E$-$8$;
$\Bar{q} = 3$. 
For SQP the following settings were passed to \texttt{fmincon}:
 \texttt{Algorithm} $=$ \texttt{sqp}; 
 fixed penalty parameter $q=1$; 
 no moving limits; 
 \texttt{TolFun} $= 1E$-$8$; 
 \texttt{TolCon} $= 1E$-$8$; 
 \texttt{MaxIter} $= 1000$; 
 \texttt{Hessian}  $=$ \texttt{bfgs}.
For GA the following settings were used:
 \texttt{PopulationSize} $=5000$;
 \texttt{StallGenLimit} $=100$;
 \texttt{TolFun} $=1E$-$8$;
 \texttt{Generations} $=1000$; 
 \texttt{UseParallel} $=$ \texttt{true}.

Lastly, all the numerical experiments with MMA and SQP were run with a single core on a standard laptop with Windows $10$, $16$ Gb of RAM, and an Intel(R) Core(TM) i7-8665U CPU running at $1.90$ GHz.
The numerical experiments with GA were run in parallel with $12$ cores on a Linux server, with $256$ Gb of RAM, and an Intel(R) Xeon(R) E5-2670 v3 CPU running at $2.30$ GHz


\subsection{Circular wind farm with radius $1300$ m}
\label{sec:ex1}
\begin{figure}[h]
\centering
\includegraphics[trim=0 0 0 0,clip,width=.85\linewidth]{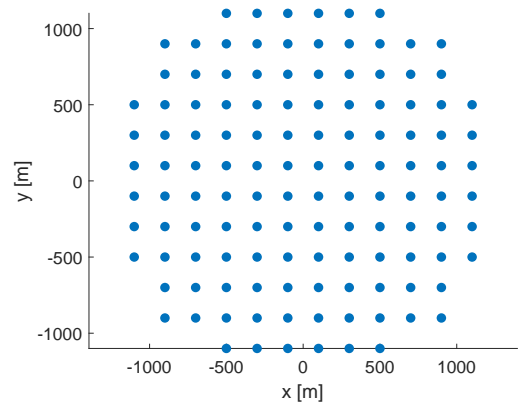}
\caption{Wind farm reference grid considered in Sec.~\ref{sec:ex1} with $124$ potential wind turbines}
\label{fig:groundstruct1}
\end{figure}

In this example we consider a circular wind farm with radius $R=1300$ m and $124$ potential wind turbines. The spacing between the potential wind turbines in the reference grid is $200$ m. The reference grid is shown in Fig.~\ref{fig:groundstruct1}.
We allow a minimum of $16$ and a maximum of $64$ wind turbines to be placed in the wind farm. 
We also impose a minimum distance of $2D\,=\,260$ m between the turbines by means of the constraints defined in Eq.~\eqref{eq:locvolcon1}. 
We consider the  IEA37 $3.4$ MW reference turbine \cite{bortolotti2019iea}, whose characteristics are listed in Tab.~\ref{tab:3.4turb}.

\begin{table}[h]
\centering
\begin{tabular}{rrl}
\toprule
Rotor diameter &        $130$ &          m \\

Turbine rating &       $3.37$ &         MW \\

Cut-in wind speed &          $4$ &        m/s \\

Rated wind speed &        $9.8$ &        m/s \\

Cut-out wind speed &         $25$ &        m/s \\
Thrust coefficient $C_T$ &         $8/9$ &        - \\
\bottomrule
\end{tabular}  
\caption{\label{tab:3.4turb}IEA37 $3.4$ MW reference wind turbine data}
\end{table}

\begin{figure}[h]
\centering
\includegraphics[trim=50 20 50 20,clip,width=.75\linewidth]{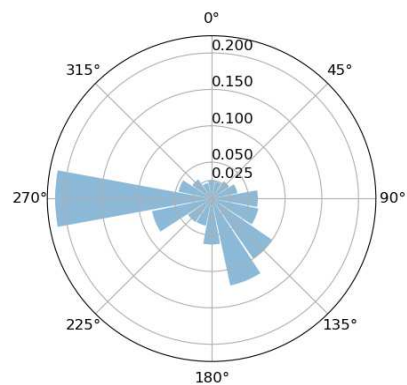}
\caption{Wind rose for a wind speed of $9.8\, m/s$}
\label{fig:windrose}
\end{figure}

We consider the wind rose from \cite{baker2019best} which is available at \cite{IEA37WFLO}. It is characterized by a constant wind speed of 9.8 m/s. The wind rose is defined by $16$ discrete bins as shown in Fig.~\ref{fig:windrose}.

\begin{figure*}[h]
\begin{subfigure}{.32\textwidth}
\centering
\includegraphics[trim=0 0 0 0,clip,width=.99\linewidth]{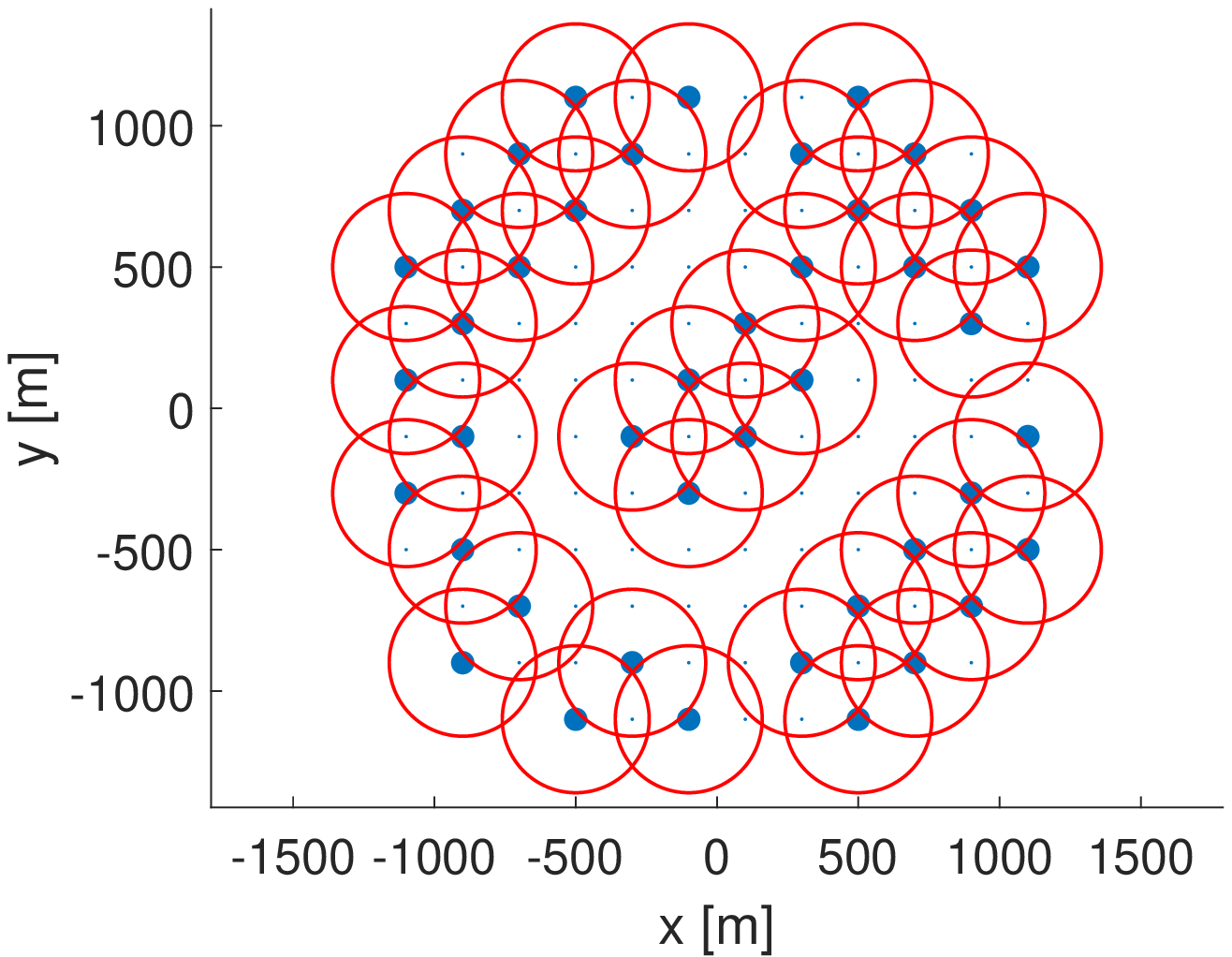}
\caption{\label{fig:optlaymma}MMA: $42$ wind turbines}
\end{subfigure}\hfill
\begin{subfigure}{.32\textwidth}
\centering
\includegraphics[trim=0 0 0 0,clip,width=.99\linewidth]{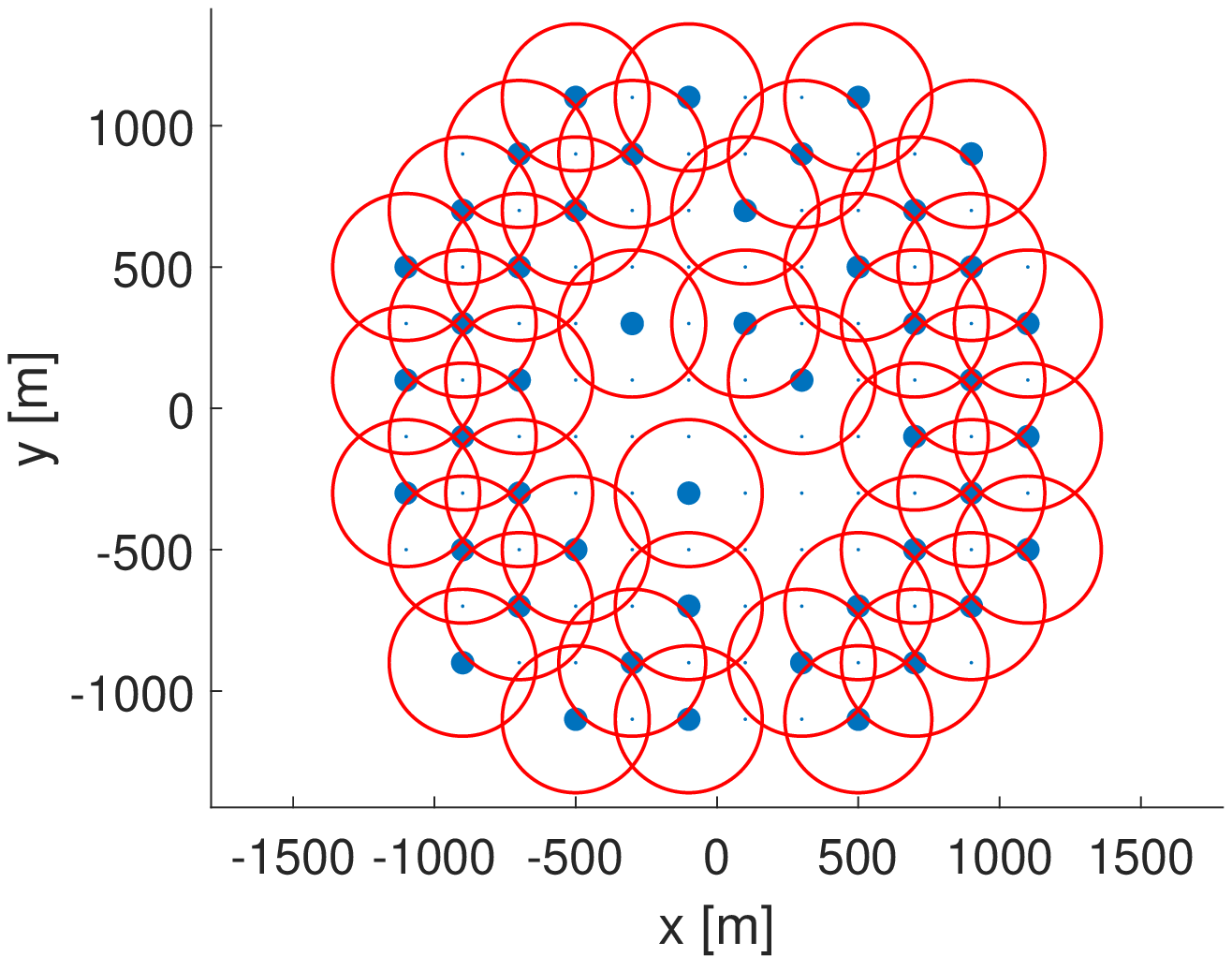}
\caption{\label{fig:optlayip}SQP: $46$ wind turbines}
\end{subfigure}\hfill
\begin{subfigure}{.32\textwidth}
\centering
\includegraphics[trim=0 0 0 0,clip,width=.99\linewidth]{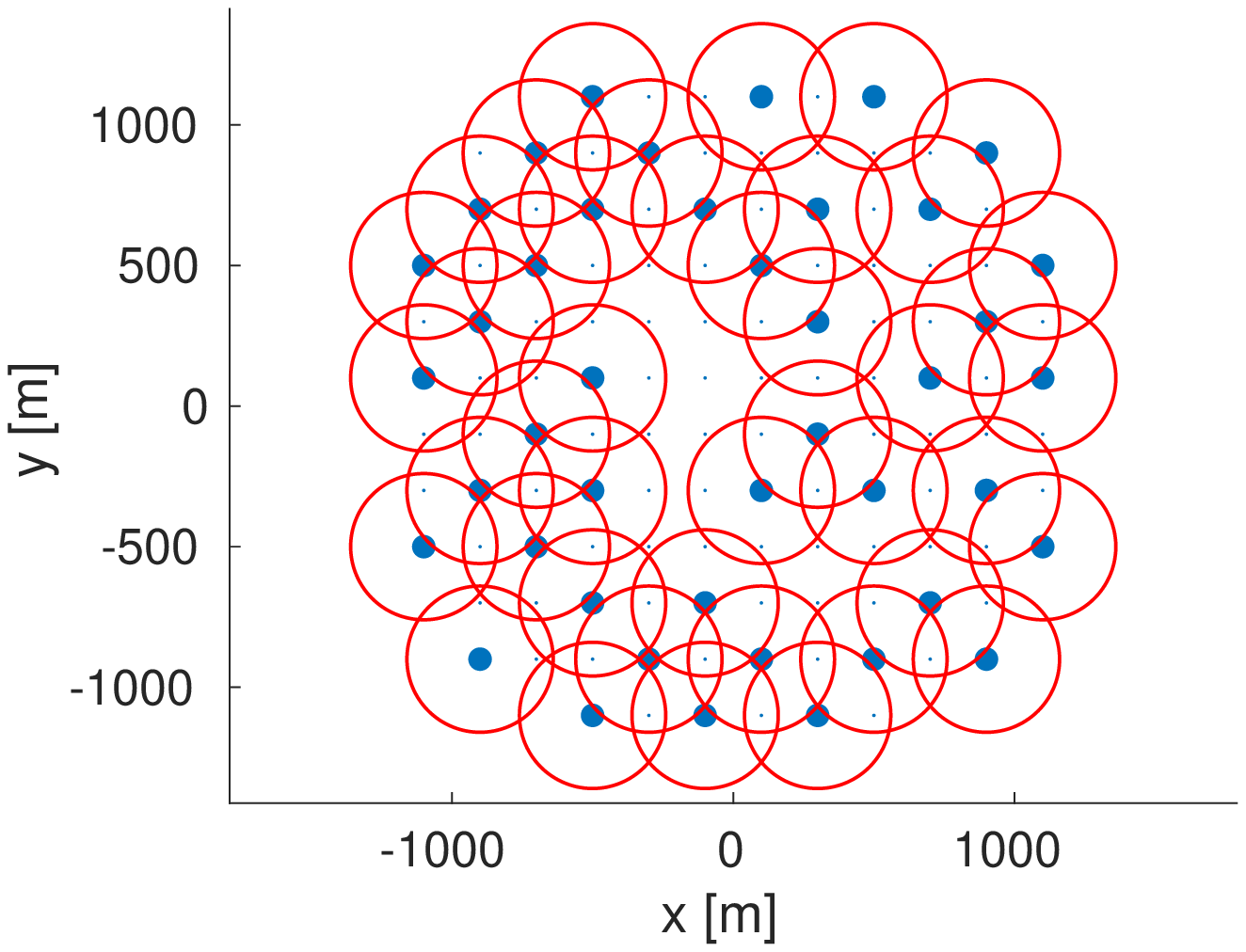}
\caption{\label{fig:optlayga}GA: $43$ wind turbines}
\end{subfigure}
\caption{\label{fig:optlayouts}Optimized wind farm layouts for the example of Sec.~\ref{sec:ex1} }
\end{figure*}

\begin{figure*}[h]
\begin{subfigure}{.32\textwidth}
\centering
\includegraphics[trim=0 0 0 0,clip,width=.99\linewidth]{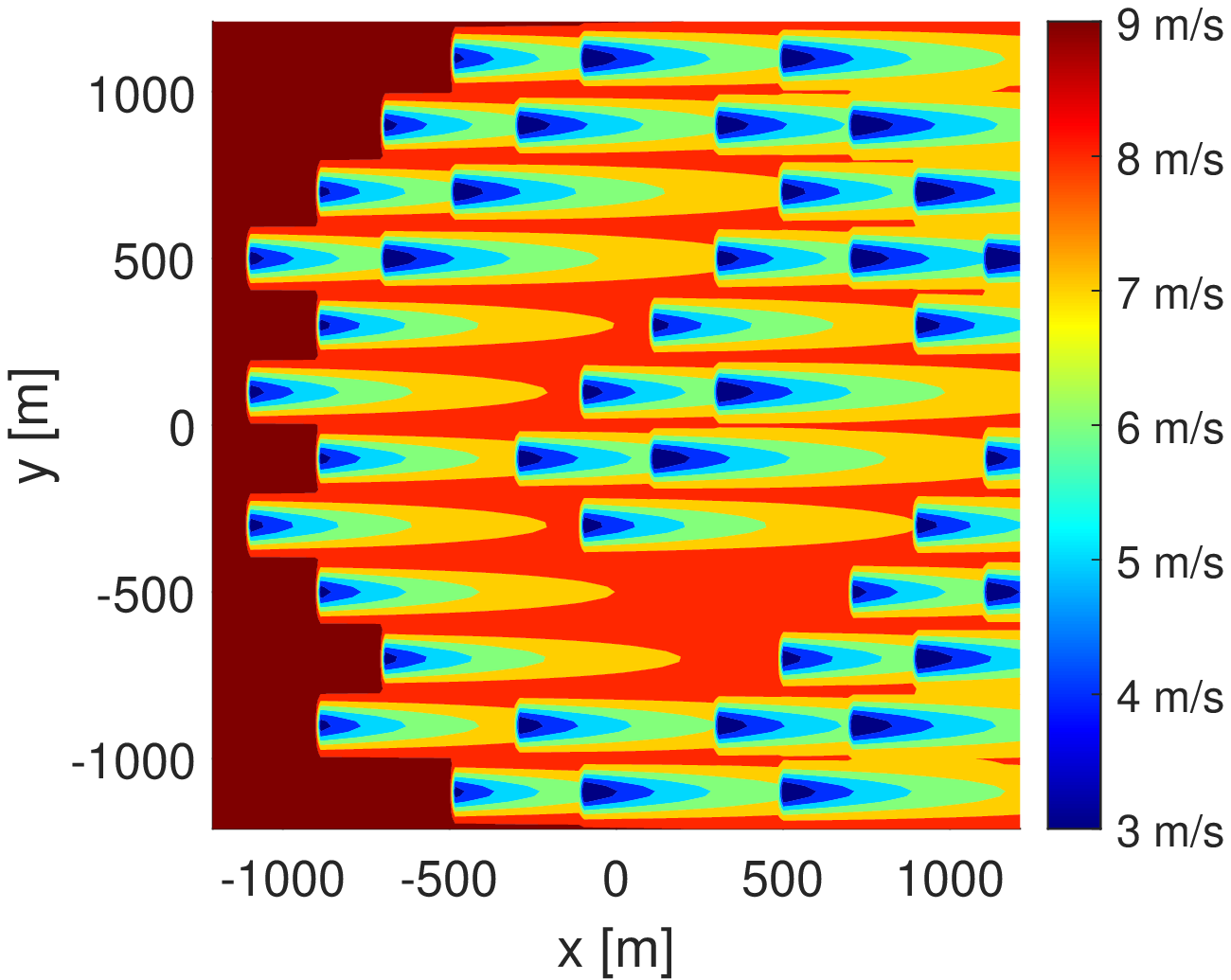}
\caption{MMA: $AEP\,=\,580.638$ GWh}
\end{subfigure}\hfill
\begin{subfigure}{.32\textwidth}
\centering
\includegraphics[trim=0 0 0 0,clip,width=.99\linewidth]{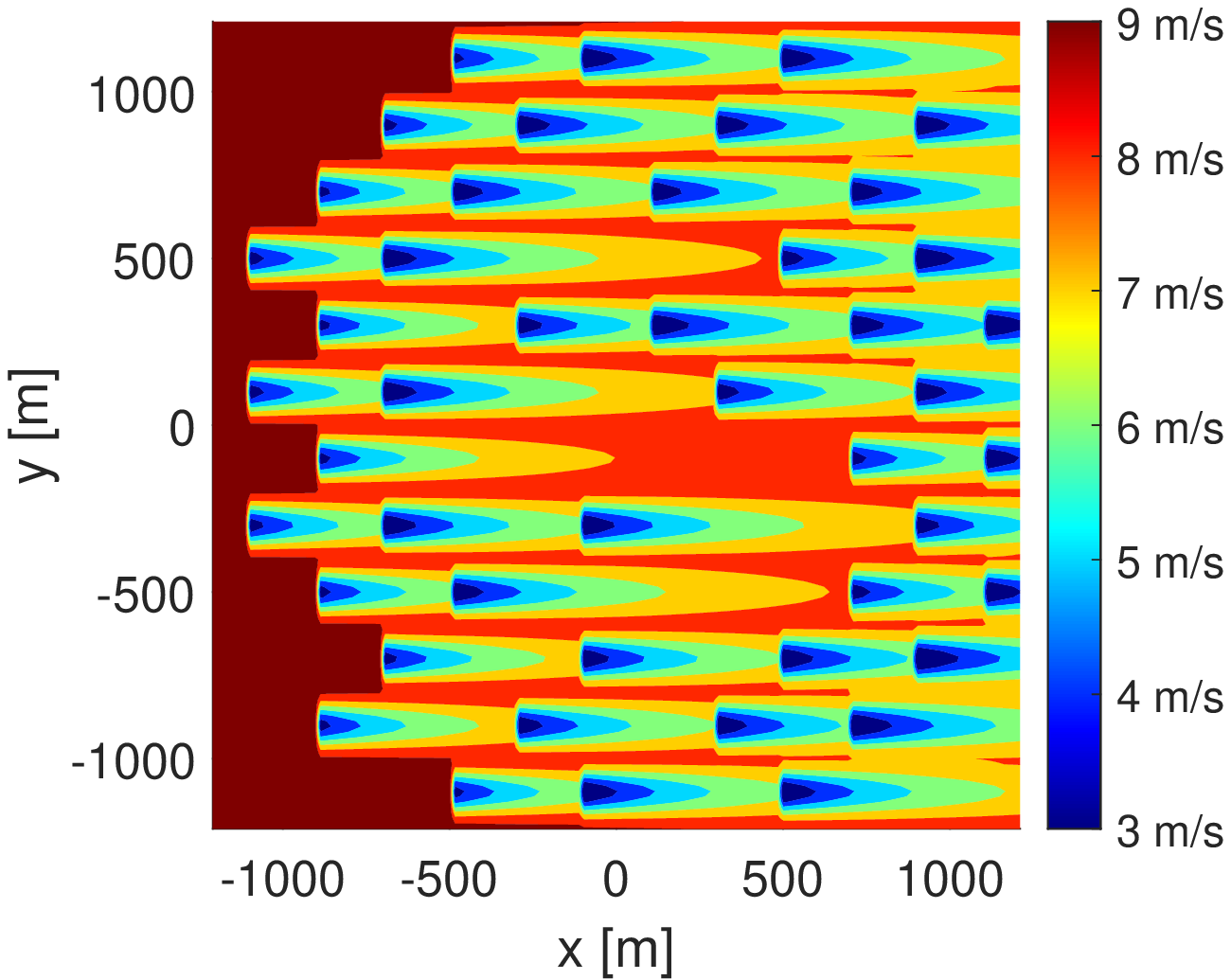}
\caption{SQP: $AEP\,=\,586.902$ GWh}
\end{subfigure}\hfill
\begin{subfigure}{.32\textwidth}
\centering
\includegraphics[trim=0 0 0 0,clip,width=.99\linewidth]{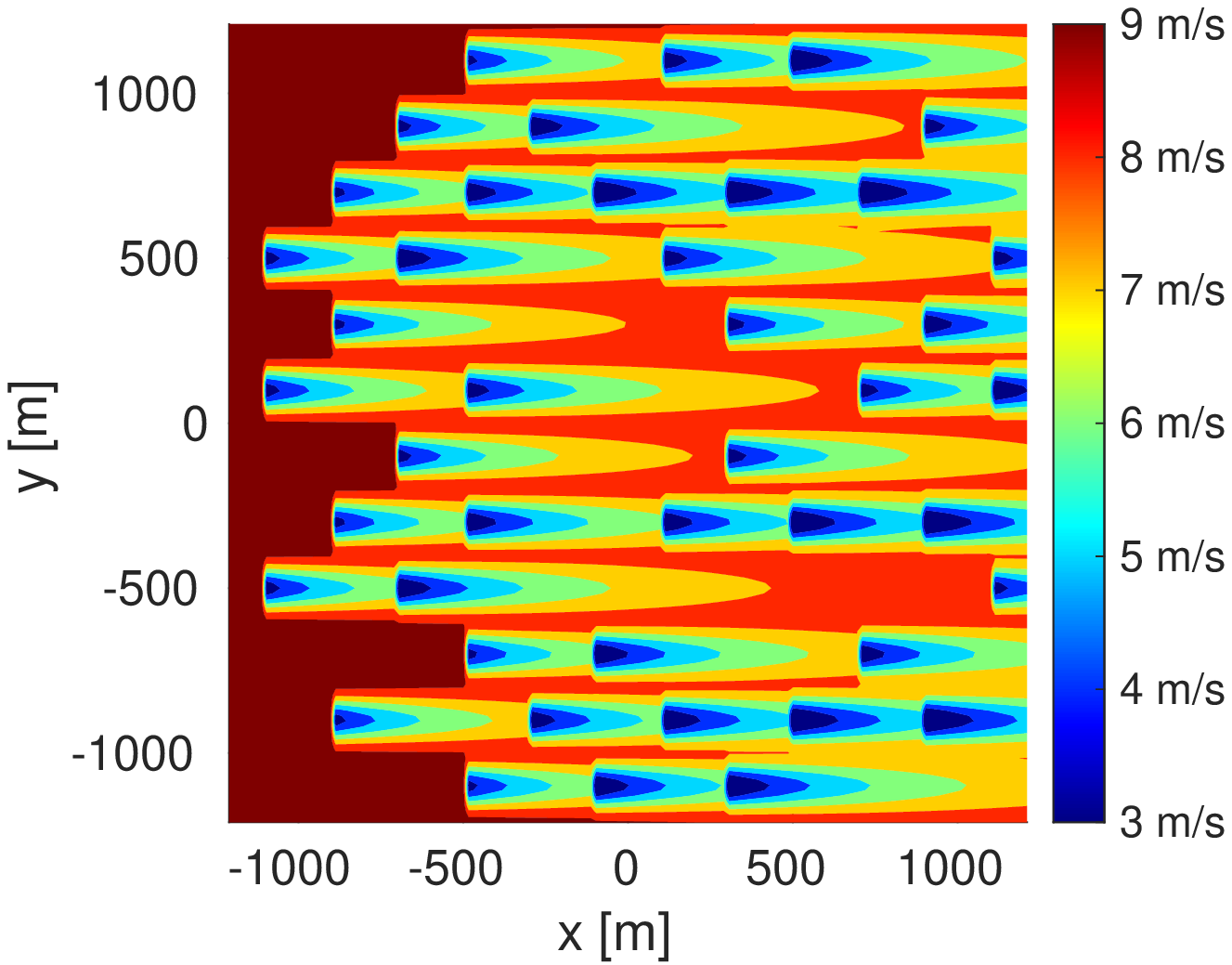}
\caption{GA: $AEP\,=\,575.584$ GWh}
\end{subfigure}
\caption{\label{fig:optlayouts2}Optimized wind farm layouts and contour of wind speed across the wind farms for the example of Sec.~\ref{sec:ex1}. The incoming wind speed direction angle is $270$ deg}
\end{figure*}

With both MMA and SQP the variables $\rho$ where initialized to $0.2$, in order to start from a feasible design point. 
The MMA algorithm converged after $72$ iterations in $5$ s, placing $42$ wind turbines. The final value of the penalty parameter was $q\,=\,3.5$, and the optimized AEP value of the wind farm was $580.638$ GWh. 
The SQP algorithm converged after $8$ iterations in $0.3$ s, placing $46$ wind turbines. The final AEP value of the wind farm was $586.902$ GWh. 
The GA algorithm converged after $209$ iterations in $9$ min $13$ s, placing $43$ wind turbines. The final AEP value of the wind farm was $575.584$ GWh. The optimized layouts are shown in Fig.~\ref{fig:optlayouts}, where the red circles mark the minimum spacing between wind turbines.
Fig.~\ref{fig:optlayouts2} shows a plot of the wind speed across the optimized wind farm layouts, for an incoming wind with a direction angle of $270$ deg.

\begin{figure}[h]
\centering
\includegraphics[trim=0 0 0 0,clip,width=.99\linewidth]{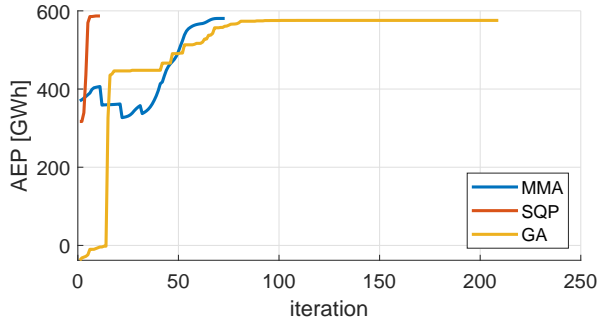}
\caption{Objective function histories of the optimization algorithms for the example of Sec.~\ref{sec:ex1} with $124$ potential wind turbines}
\label{fig:conhist1}
\end{figure}

Fig.~\ref{fig:conhist1} shows the evolution of AEP during the optimization analyses with the different algorithms. 
MMA and SQP converge quickly to the final solutions, whereas GA requires several initial iterations (associated with negative values of AEP) before it identifies feasible design individuals. Moreover, GA requires far more iterations and converges to a layout with the lowest AEP. 

The numerical results of the example of Sec.~\ref{sec:ex1} are summarized in Table ~\ref{tab:comparison1} in order to allow for a comparison of the different optimization algorithms in terms of final design and computational effort invested. 
From Table ~\ref{tab:comparison1}, it is possible to first observe that in this example SQP identified the best wind farm layout in terms of final AEP. 
SQP required also the least computational time and effort. MMA converged to a wind farm layout with a final AEP $1.1$ $\%$ lower than that of SQP. 
In general, GA is an algorithm known for being less likely affected by the presence of local minima, and in some problems it may converge towards the global optima (even though there is not a formal guarantee for that). 
However, in this case it converged to the worst layout compared to MMA and SQP, probably because of the high combinatorial nature of the problem at hand.
Moreover, it required a computational time significantly higher than MMA and SQP ($1.1$ $10^2$ times higher than MMA and $1.8$ $10^3$ times higher than SQP, using different hardware resources).
\begin{table*}[h]
\centering
\begin{tabular}{ccccccc}
\toprule
           &  Number of &        AEP & Iterations &   AEP function & Computational &   Hardware \\

           &   turbines &            &            & evaluations &       time &            \\
\midrule
       MMA &         42 & 580.638 GWh &         72 &         72 &      5 s &     1 core @ 1.9 GHz \\

       SQP &         46 & 586.902 GWh &          8 &          17 &      0.3 s &     1 core @ 1.9 GHz \\

        GA &         43 & 575.584 GWh &        209 &     1\,048\,960 & 9 min 13 s &   12 cores @ 2.3 GHz \\
\bottomrule
\end{tabular}  
\caption{\label{tab:comparison1}Comparison of the results obtained with the different optimization algorithms in Sec.~\ref{sec:ex1}. MMA and SQP solved problem \eqref{eq:endoptprob}, whereas GA solved problem \eqref{eq:initoptprob}}
\end{table*}


\subsection{Circular wind farm with radius $3000$ m}
\label{sec:ex2}
\begin{figure}[h]
\centering
\includegraphics[trim=0 0 0 0,clip,width=.85\linewidth]{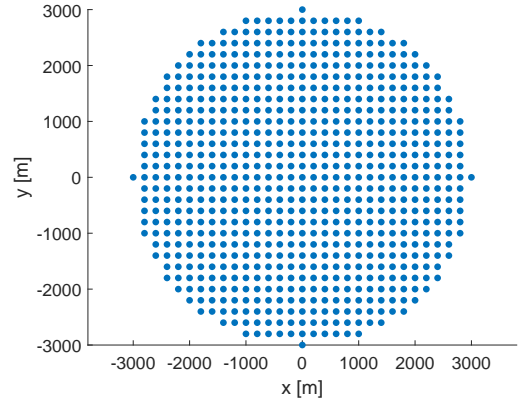}
\caption{Wind farm reference grid considered in Sec.~\ref{sec:ex2} with $709$ potential wind turbines}
\label{fig:groundstruct2}
\end{figure}

We now consider a circular wind farm with radius $R=3000$ m and $709$ potential wind turbines. The spacing between the potential wind turbines in the reference grid is also in this case $200$ m. The reference wind farm grid is shown in Fig.~\ref{fig:groundstruct2}.
We allow a minimum of $64$ and a maximum of $256$ wind turbines to be placed in the wind farm. 
The remaining settings are the same as in Sec.~\ref{sec:ex1}.

\begin{figure*}[h]
\begin{subfigure}{.32\textwidth}
\centering
\includegraphics[trim=0 0 0 0,clip,width=.99\linewidth]{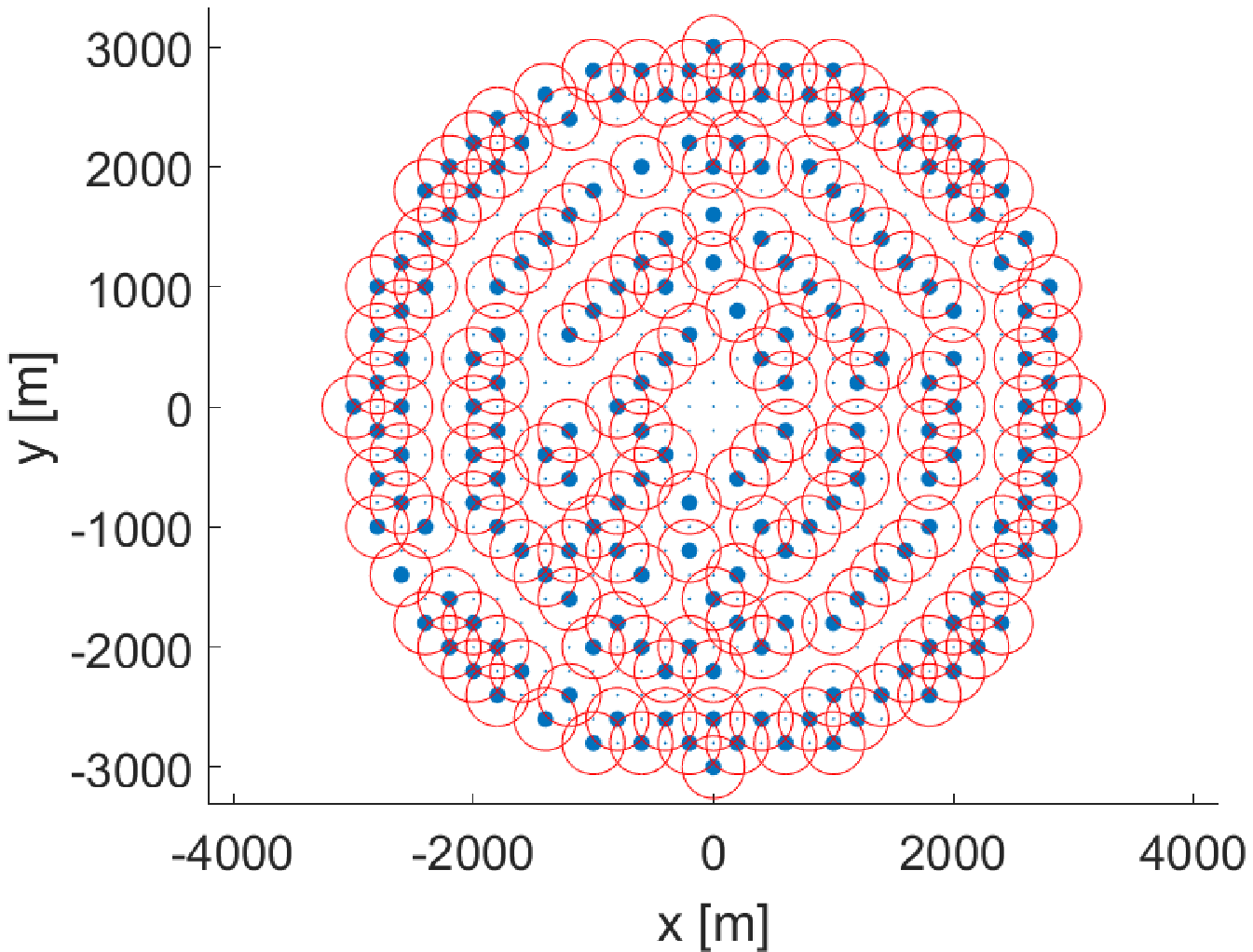}
\caption{\label{fig:optlaymma}MMA: $197$ wind turbines}
\end{subfigure}\hfill
\begin{subfigure}{.32\textwidth}
\centering
\includegraphics[trim=0 0 0 0,clip,width=.99\linewidth]{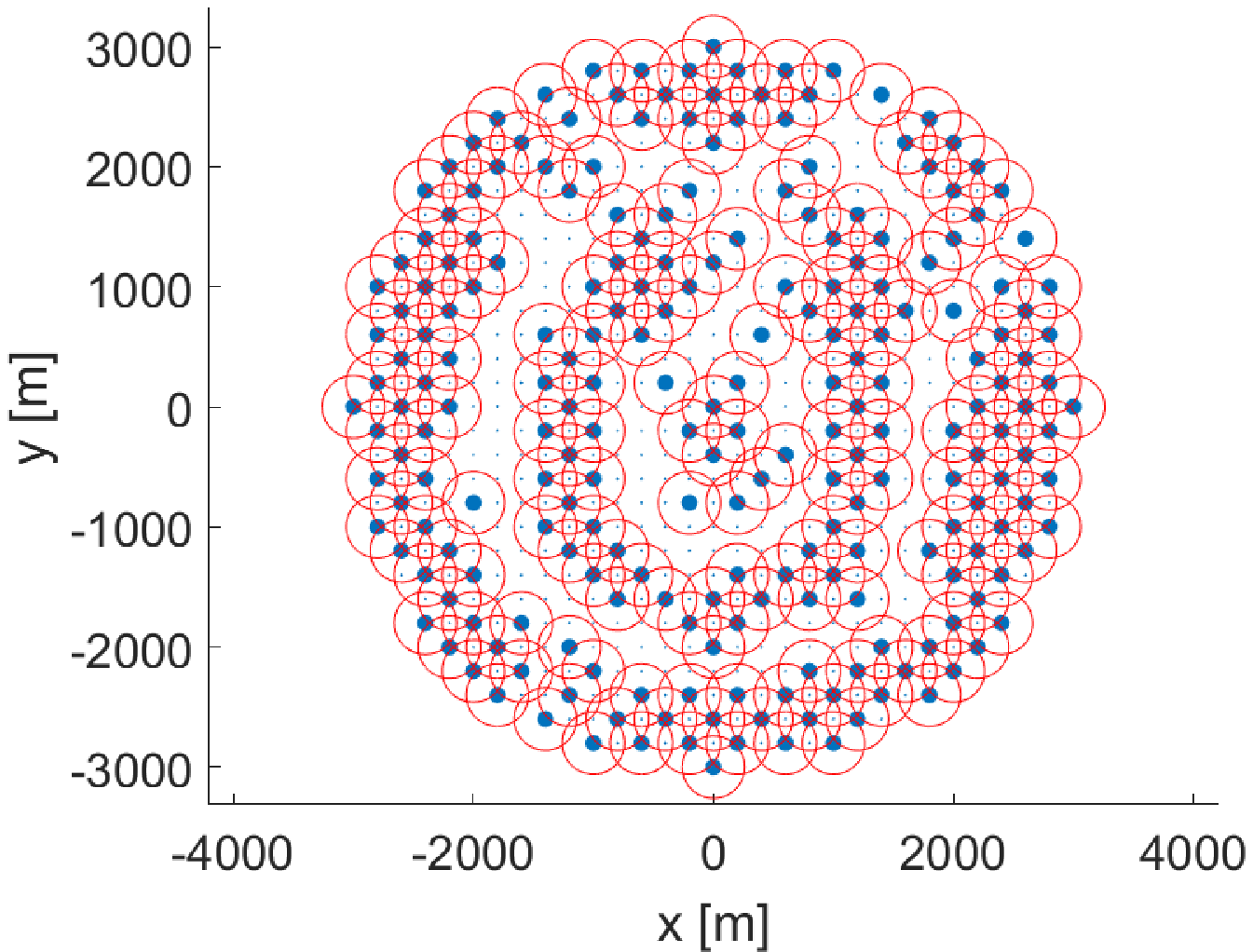}
\caption{\label{fig:optlayip}SQP: $231$ wind turbines}
\end{subfigure}\hfill
\begin{subfigure}{.32\textwidth}
\centering
\includegraphics[trim=0 0 0 0,clip,width=.99\linewidth]{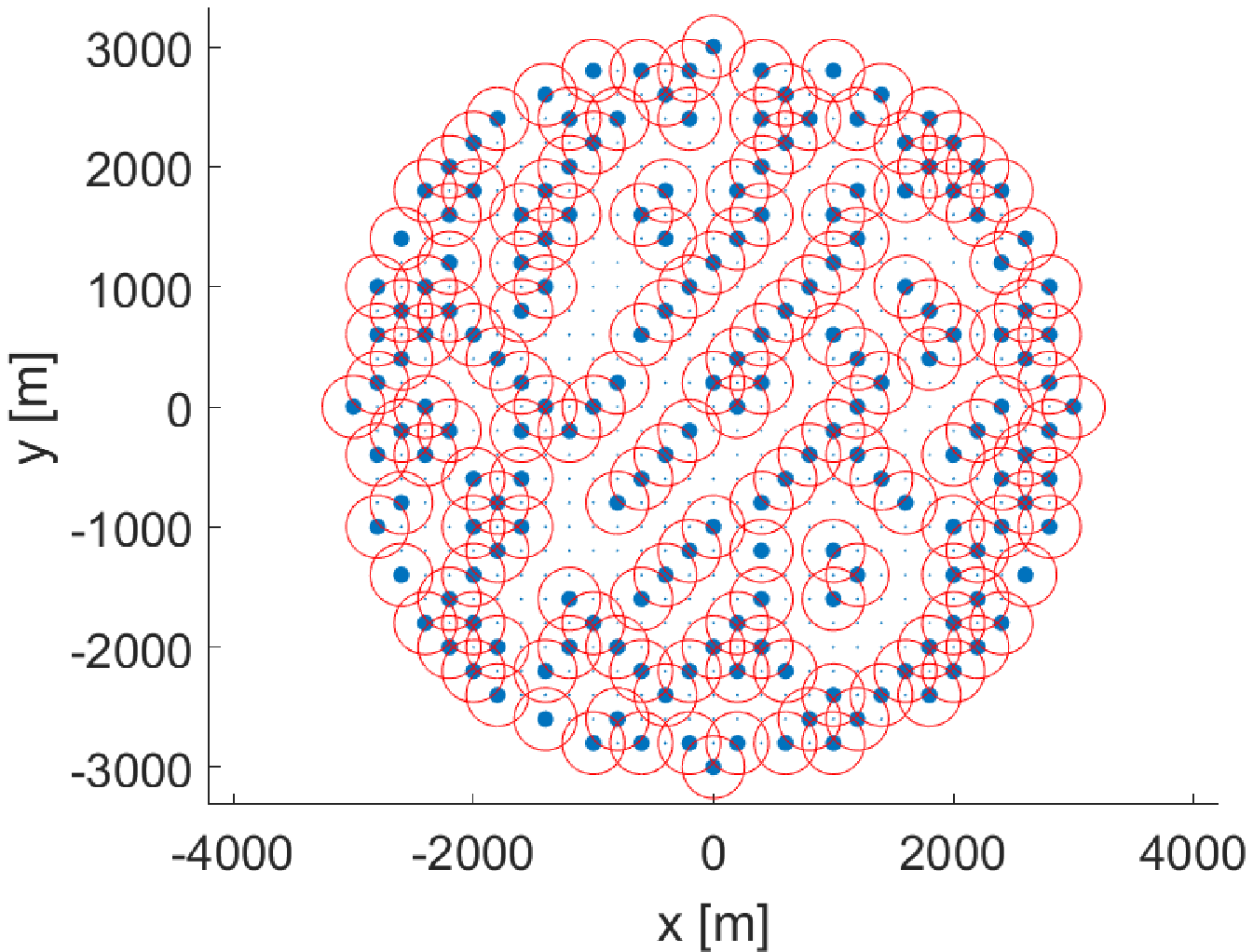}
\caption{\label{fig:optlayga}GA: $189$ wind turbines}
\end{subfigure}
\caption{\label{fig:optlayouts3}Optimized wind farm layouts for the example of Sec.~\ref{sec:ex2} }
\end{figure*}

With both MMA and SQP the variables $\rho$ where initialized to $0.1805$, in order to start from a feasible design point. 
The MMA algorithm converged after $83$ iterations in $8$ min $47$ s, placing $197$ wind turbines. 
The final value of the penalty parameter was $q\,=\,4$, and the optimized AEP value of the wind farm was $2190.576$ GWh. 
The SQP algorithm converged after $8$ iterations in $27$ s, placing $231$ wind turbines. The final AEP value of the wind farm was $2088.646$ GWh. 
The population size for GA was increased to $10\,000$ individuals in this example (i.e. \texttt{PopulationSize} $=10000$). The GA algorithm converged after $375$ iterations in $35$ h $35$ min $0$ s, placing $189$ wind turbines. The final AEP value of the wind farm was $2181.170$ GWh. The optimized layouts are shown in Fig.~\ref{fig:optlayouts3}, where the red circles mark the minimum spacing between wind turbines.

\begin{figure*}[h]
\begin{subfigure}{.32\textwidth}
\centering
\includegraphics[trim=0 0 0 0,clip,width=.99\linewidth]{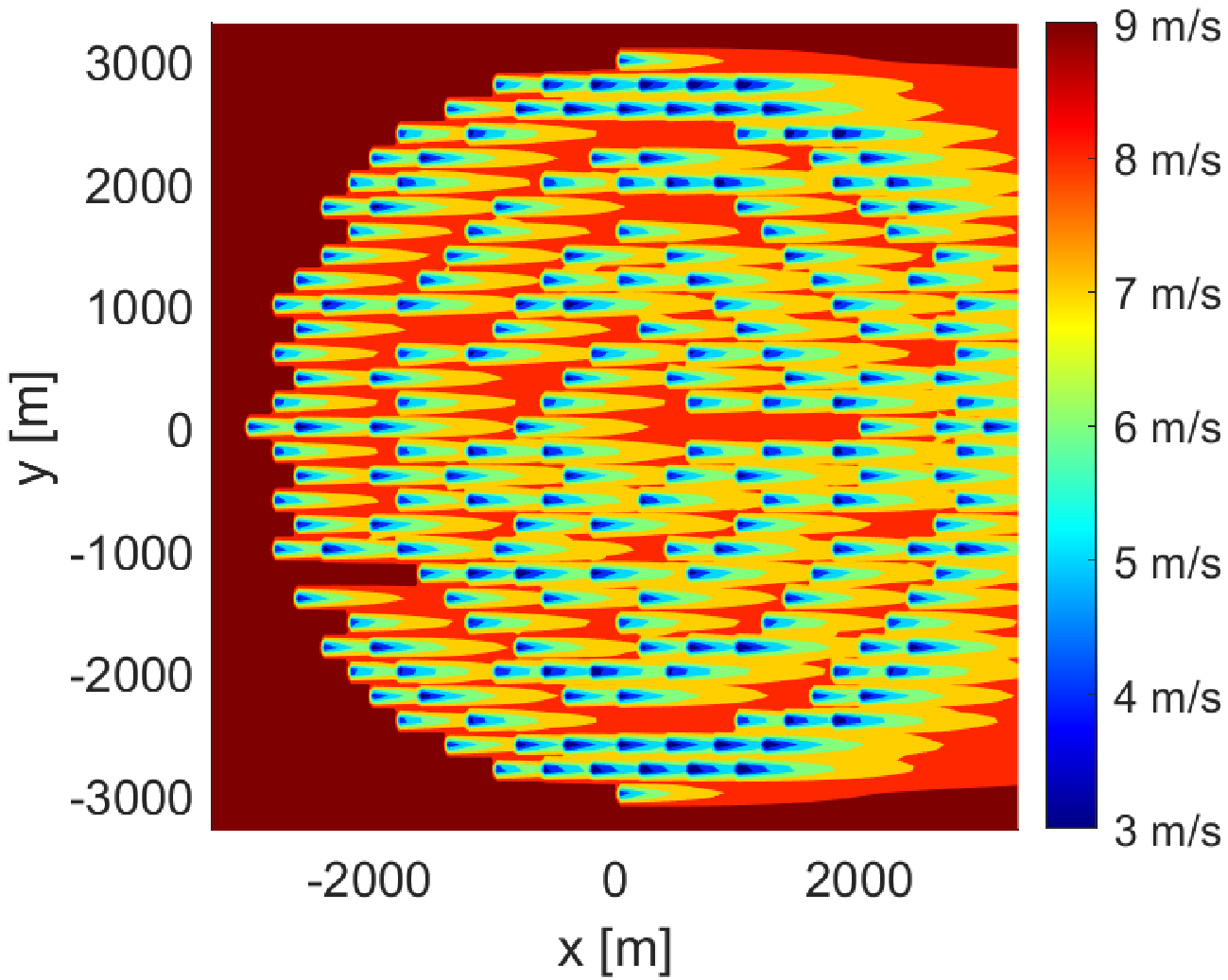}
\caption{MMA: $AEP\,=\,2190.576$ GWh}
\end{subfigure}\hfill
\begin{subfigure}{.32\textwidth}
\centering
\includegraphics[trim=0 0 0 0,clip,width=.99\linewidth]{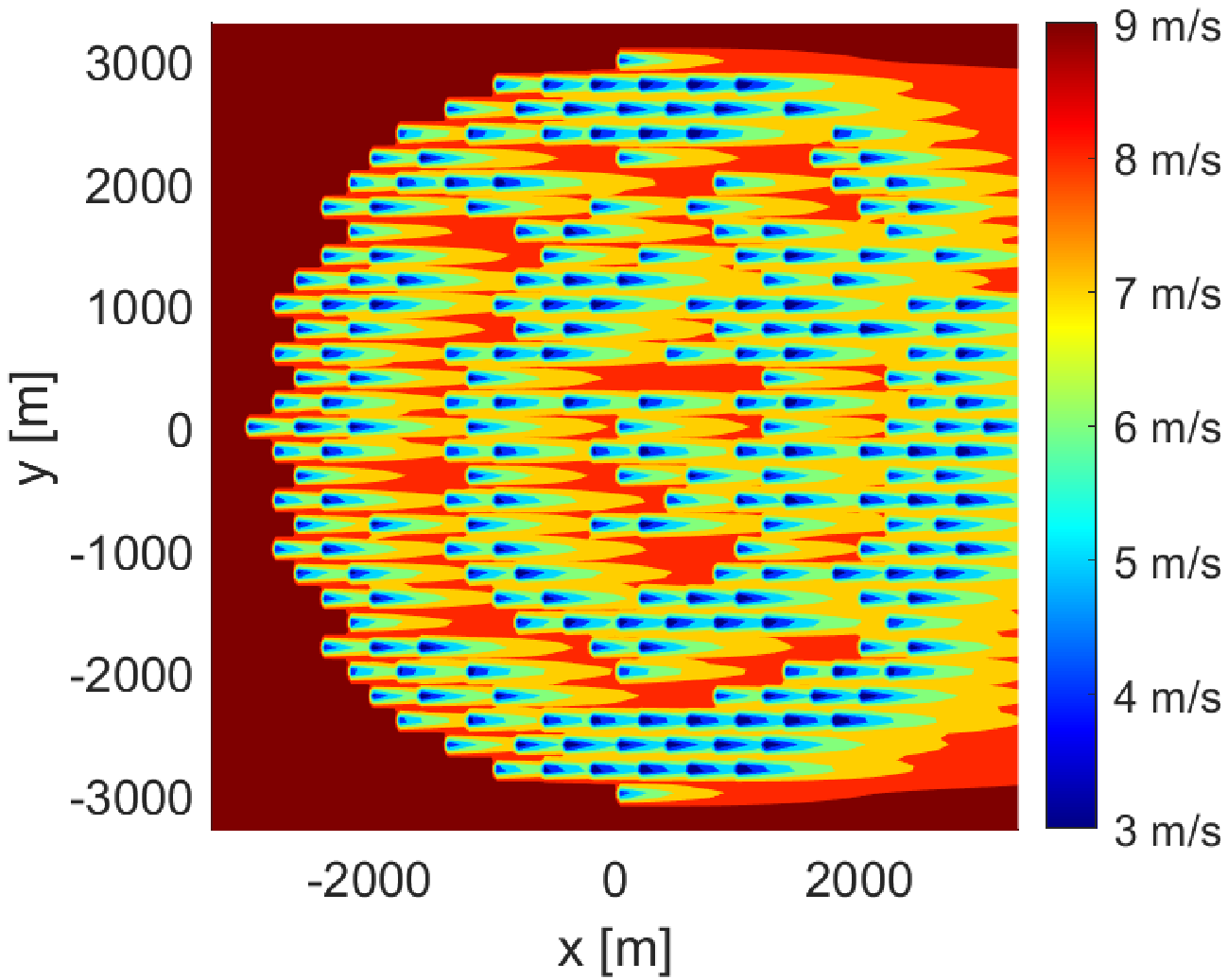}
\caption{SQP: $AEP\,=\,2088.646$ GWh}
\end{subfigure}\hfill
\begin{subfigure}{.32\textwidth}
\centering
\includegraphics[trim=0 0 0 0,clip,width=.99\linewidth]{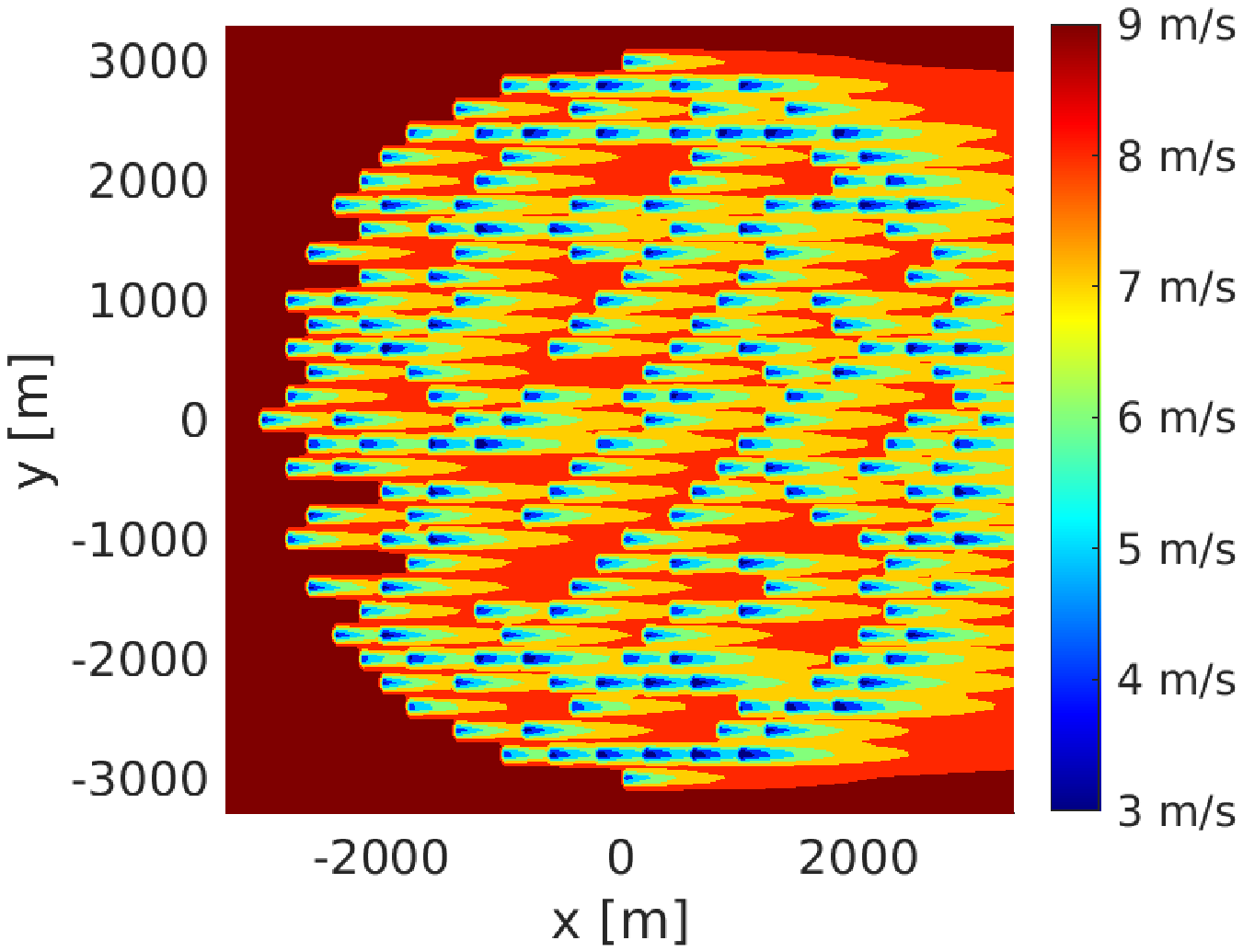}
\caption{GA: $AEP\,=\,2181.170$ GWh}
\end{subfigure}
\caption{\label{fig:optlayouts4}Optimized wind farm layouts and contour of wind speed across the wind farms for the example of Sec.~\ref{sec:ex2}. The incoming wind speed direction is $270$ deg}
\end{figure*}

Fig.~\ref{fig:optlayouts4} shows a plot of the wind speed across the optimized wind farm layouts, for an incoming wind with an angle of $270$ deg.

\begin{figure}[h]
\centering
\includegraphics[trim=0 0 0 0,clip,width=.99\linewidth]{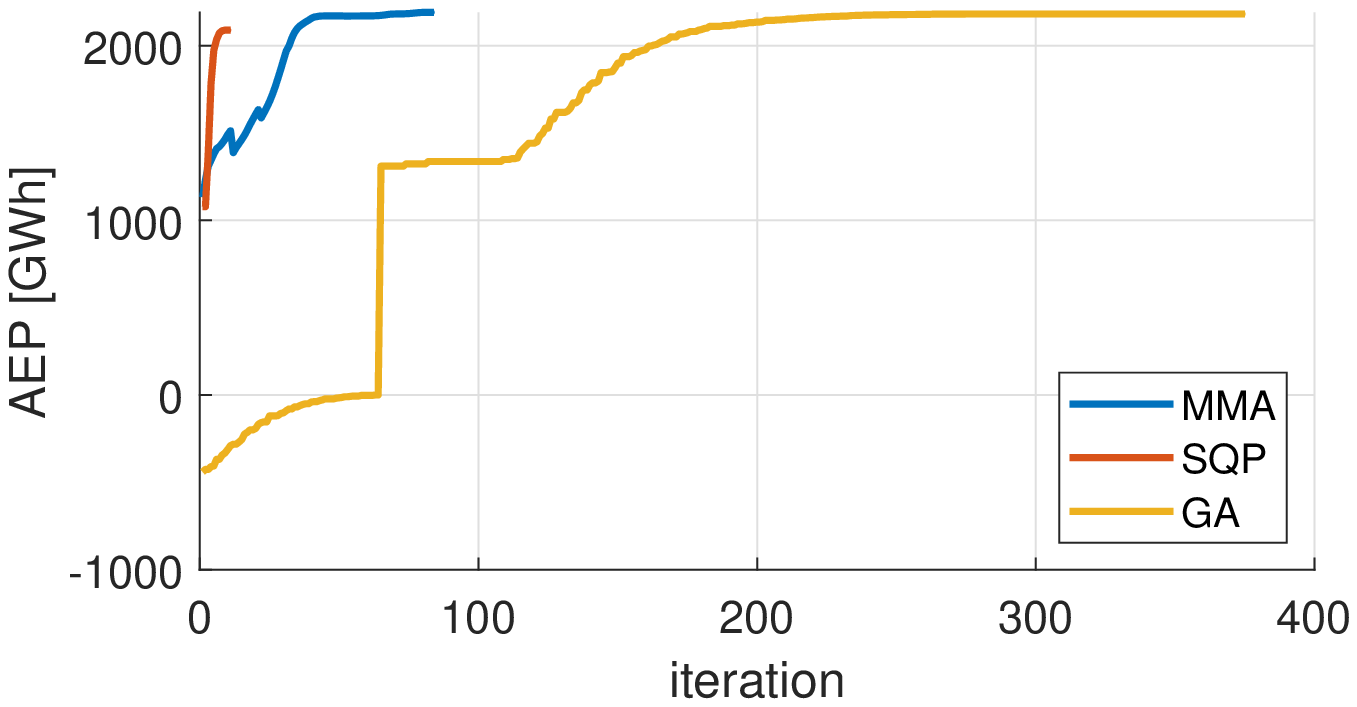}
\caption{Objective function histories of the optimization algorithms for the example of Sec.~\ref{sec:ex2} with $709$ potential wind turbines}
\label{fig:conhist2}
\end{figure}

Fig.~\ref{fig:conhist2} shows the evolution of AEP during the optimization analyses with the different algorithms. 
As expected, also in this case MMA and SQP converged more quickly to the final layouts. MMA identified the best final layout in terms of AEP compared to SQP and GA.

The numerical results of the example of Sec.~\ref{sec:ex1} are summarized in Table ~\ref{tab:comparison2}, in order to compare the final solutions attained. 
From Table ~\ref{tab:comparison2} it is possible to first observe that MMA converged to the best final layout in terms of AEP. 
SQP required the least computational time and effort, but converged to a wind farm layout with a final AEP $4.7$ $\%$ lower than that of MMA. 
GA required the highest computational time to converge to the final layout ($2.43$ $10^2$ times higher than MMA and $4.75$ $10^3$ times higher than SQP, using different hardware resources).
\begin{table*}[h]
\centering
\begin{tabular}{ccccccc}
\toprule
           &  Number of &        AEP & Iterations &   Function & Computational &   Hardware \\

           &   turbines &            &            & evaluations &       time &            \\
\midrule
       MMA &        $197$ & $2190.576$ GWh &         $83$ &         $83$ &       $8$ min $47$ s & $1$ core @ $1.9$ GHz \\

       SQP &        $231$ & $2088.646$ GWh &          $8$ &         $17$ &       $27$ s & $1$ core @ $1.9$ GHz \\

        GA &        $189$ & $2181.170$ GWh &        $375$ &    $3\,758\,130$ & $35$ h $35$ min $00$ s & $12$ cores @ $2.3$ GHz \\
\bottomrule
\end{tabular}  
\caption{\label{tab:comparison2}Comparison of the results obtained with the different optimization algorithms in Sec.~\ref{sec:ex2}. MMA and SQP solved problem \eqref{eq:endoptprob}, whereas GA solved problem \eqref{eq:initoptprob}}
\end{table*}\


\subsection{Influence of the interpolation scheme}
\label{sec:ex3}

We run additional optimization simulations with MMA and SQP for the examples of Sec.~\ref{sec:ex1} and Sec.~\ref{sec:ex2}. 
This time we considered a linear interpolation of the design variables $\rho$, by setting the penalty parameter $q$ to zero. 
The purpose of these additional analyses was to explore the effect of the RAMP interpolation scheme onto the final optimized wind farm layouts.
In Table~\ref{tab:comparison3} and Table~\ref{tab:comparison4} the results in terms of final AEP and number of wind turbines placed are summarized.

In the case of the example of Sec.~\ref{sec:ex1}, both MMA and SQP converged to a layout with worse final AEP compared to the case with a penalty parameter $q$ different from zero.
In the case of the example of Sec.~\ref{sec:ex2} instead, the fact of not using the RAMP scheme led to better final results with MMA, and worse final results with SQP in terms of final AEP.

\begin{figure}[h]
\centering
\includegraphics[trim=0 0 0 0,clip,width=.99\linewidth]{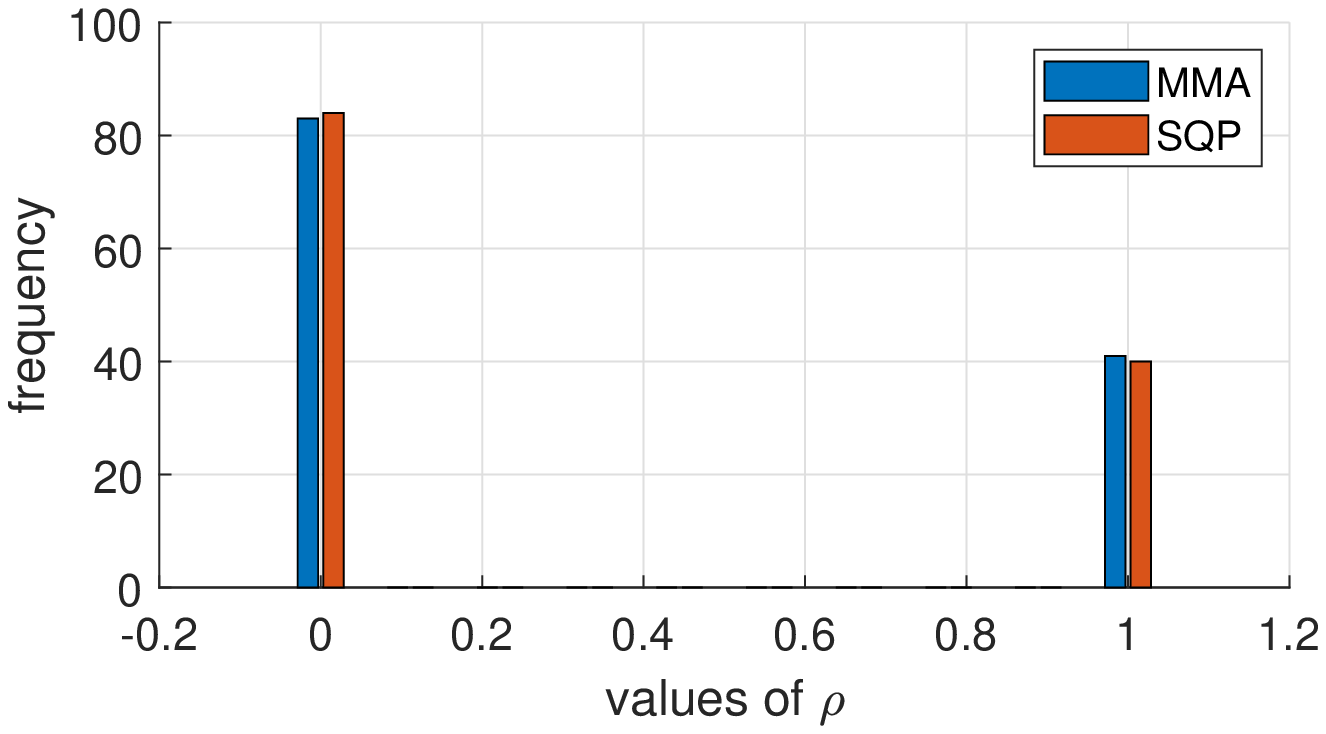}
\caption{Histogram of the optimized values of the variables $\rho$ obtained with MMA and SQP without RAMP for the example of Sec.~\ref{sec:ex1} with $124$ potential wind turbines}
\label{fig:histog1}
\end{figure}

\begin{figure}[h]
\centering
\includegraphics[trim=0 0 0 0,clip,width=.99\linewidth]{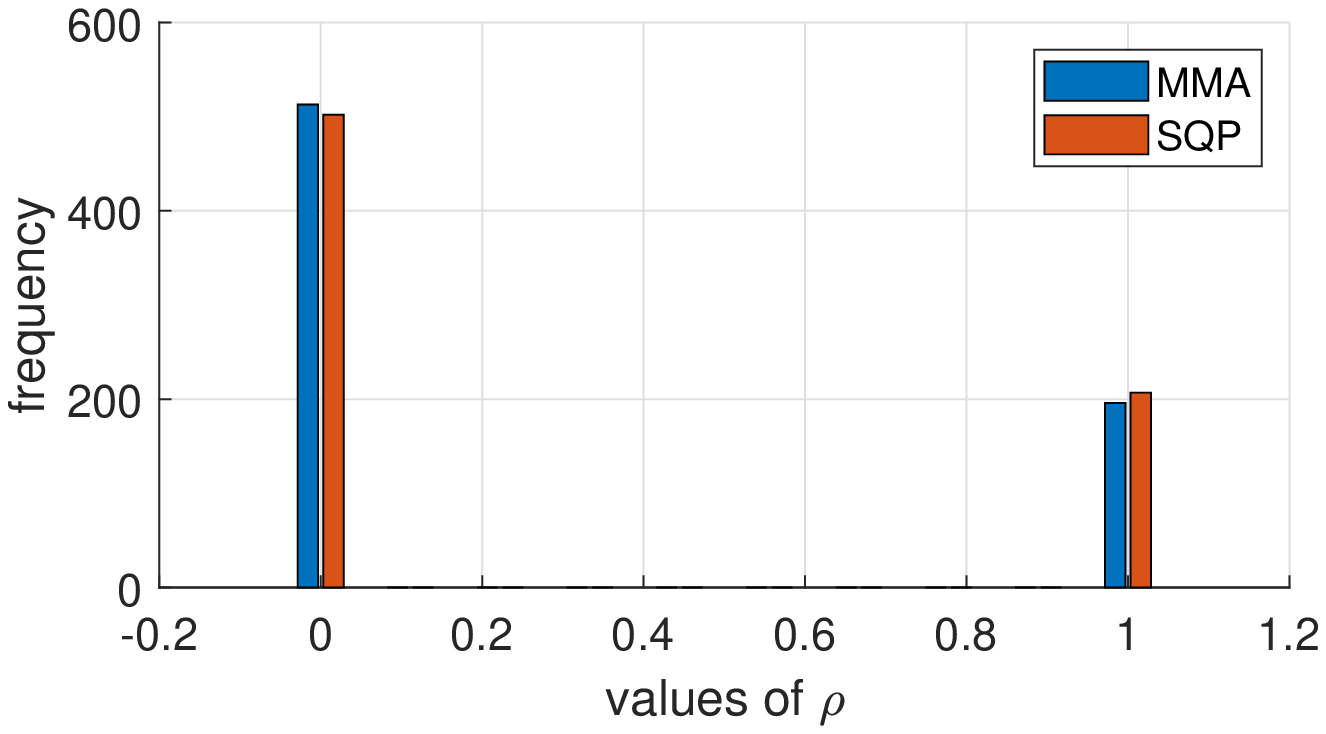}
\caption{Histogram of the optimized values of the variables $\rho$ obtained with MMA and SQP without RAMP for the example of Sec.~\ref{sec:ex2} with $709$ potential wind turbines}
\label{fig:histog2}
\end{figure}

Fig.~\ref{fig:histog1} and Fig.~\ref{fig:histog2} show the histograms of the final optimized values of the variables $\rho$ obtained with MMA and SQP without the RAMP interpolation scheme in the two numerical examples of Sec.~\ref{sec:ex1} and Sec.~\ref{sec:ex2}. 
It can be observed that at the end of the optimization, the final values of the optimization variables tend towards crisp $0$-$1$ distributions of their values.
This is quite surprising, as in general the use of material interpolation scheme in the field of topology optimization has been essential in order to converge towards near discrete final designs. 
The reason for this can be explained perhaps by the proposed model, which happen to be self penalized, and it is able to promote the convergence of the optimizer towards discrete final designs even when the penalization of the RAMP functions is switched off.

It is hard to conclude whether the use of RAMP leads to better or worse designs. Different values of the parameter $q$ generate different nonconvex objective functions from and optimization perspective, and there may be certain lucky combinations of the model parameters and initial design that drive the optimization algorithms towards better local optimal solutions.
In general it is recommended to rely on the RAMP interpolation function when implementing the proposed approach. 
In numerous topology optimization problems and applications it has been observed that to successfully promote the convergence towards final near discrete designs the interpolation functions play a crucial role. It is then the responsibility of the optimization engineer to perform numerical experiments and to define the best settings for the given problem studied.

\begin{table}[h]
\centering
\begin{tabular}{lcc}
\toprule
           &  Number of &        AEP \\

           &   turbines &            \\
\midrule
MMA (w/ RAMP) &         42 & 580.638 GWh \\

MMA (w/o RAMP) &         41 & 576.592 GWh \\

SQP (w/ RAMP) &         46 & 586.902 GWh \\

SQP (w/o RAMP) &         40 & 576.486 GWh \\
 \bottomrule
\end{tabular}  
\caption{\label{tab:comparison3}Comparison of the results obtained with MMA and SQP with (w/) and without (w/o) the RAMP interpolation scheme in the example of Sec.~\ref{sec:ex1}}
\end{table}

\begin{table}[h]
\centering
\begin{tabular}{lcc}
\toprule
           &  Number of &        AEP \\

           &   turbines &            \\
\midrule
MMA (w/ RAMP) &        197 & 2190.576 GWh \\

MMA (w/o RAMP) &        196 & 2199.750 GWh \\

SQP (w/ RAMP) &        231 & 2088.646 GWh \\

SQP (w/o RAMP) &        207 & 2068.463 GWh \\
\bottomrule
\end{tabular}  
\caption{\label{tab:comparison4}Comparison of the results obtained with MMA and SQP with (w/) and without (w/o) the RAMP interpolation scheme in the example of Sec.~\ref{sec:ex2}}
\end{table}

\section{Final considerations}
\label{sec:end}

This paper presents a novel approach for the solution of the wind farm layout optimization problem.
The proposed approach relies on a density-based topology optimization approach, where continuous density variables that vary between zero and one are assigned to each potential wind turbine location.
A wind turbine exists if its associated variable equals one, otherwise it does not exist if the associated variable equals zero.
The wind farm layout optimization problem is initially formulated as a mixed-integer nonlinear programming problem. 
Its variables definitions are subsequently relaxed in order to allow for the solution of the relaxed problem with efficient gradient-based optimization algorithms. 
We use a material interpolation scheme traditionally used in the context of multi-material topology optimization to penalize the intermediate values of the design variables.
As a result of the penalization, the intermediate values of the design variables become uneconomical and the optimization algorithm is implicitly pushed towards a preference of crisp $0$-$1$ final values.

We maximize the wind farm annual energy production (AEP) with constraints on the minimum and maximum number of wind turbines placed, and on the minimum spacing between the existing wind turbines.
The optimization problem is solved with a gradient-based algorithm based on first-order information. 
The analytical gradients of the functions involved are calculated directly and provided to the optimization algorithm.

In the two numerical examples considered, the initial mixed-integer problem formulation was solved with a genetic algorithm (GA), for comparison.
The relaxed and penalized continuous problem reformulation was solved with two gradient-based algorithms: the Method of Moving Asymptotes (MMA), and a Sequential Quadratic Programming (SQP) algorithm.
The numerical results show that the gradient-based algorithms considered in this work required significantly lower computational efforts and time to identify final optimized wind farm layouts compared to GA. This is most likely due to the fact that the problem at hand is highly nonconvex and combinatorial.
SQP converged to the best layout in terms of AEP, in the first example. In the second example, the best layout was identified by MMA. 

The numerical results show that the proposed approach represents a promising interactive and easy to implement framework for solving the wind farm layout optimization problem, compared to the methods currently available in the literature. The proposed approach is expected to assist engineers in the initial conceptual wind farm design phases by allowing faster design reiterations. In a more detailed design phase, the results obtained with the proposed method could be used as starting points for subsequent more detailed high fidelity analyses and design steps. 


\section*{Acknowledgments}
The author thanks Prof. Krister Svanberg for providing the MMA MATLAB implementation. 
The author thanks also Prof. Mathias Stolpe for fruitful discussions on the topic of this paper, and for his valuable feedback on the paper draft.

\bibliography{mybibfile.bib}

\end{document}